\documentclass[sigconf]{acmart}

\usepackage{fontawesome}
\usepackage{multirow}
\definecolor{Gallery}{rgb}{0.937,0.937,0.937}
\usepackage{colortbl}
\usepackage{soul}
\usepackage{enumitem}
\usepackage{fp}
\usepackage{booktabs} 
\usepackage[T1]{fontenc}
\usepackage{tabularx}   

\definecolor{darkgreen}{rgb}{0.0, 0.6, 0.2} 
\definecolor{lightgreen}{rgb}{0.3, 0.5, 0.4} 

\AtBeginDocument{%
  }

\begin{document}

\def\barChart#1{%
  \begingroup
  \FPmul\BarLen{#1}{0.015}%
  \raisebox{0.1ex}{%
    \hbox{%
      {\color[rgb]{0.30,0.60,0.45}\rule{\BarLen cm}{1.2em}}%
      \hspace{0.4em}%
      {\normalsize\bfseries #1\%}%
    }%
  }%
  \endgroup
}



\makeatletter
\def\miniHistogram{\@ifnextchar[{\miniHistogram@opt}{\miniHistogram@opt[5]}}%

\def\miniHistogram@opt[#1]#2{%
  \leavevmode
  \vbox{%
    \hbox{%
      \def\maxval{#1}%
      \def\barwidth{6pt}
      \def\totalH{15}
      \def\do##1{%
        \FPeval\h{(##1)/(\maxval)*(\totalH)}%
        {\color{black!70}\rule{\barwidth}{\h pt}}\hspace{0.5pt}%
      }%
      \forcsvlist{\do}{#2}%
    }%
    \hrule height 0.8pt%
  }%
}
\makeatother

\title{\textit{``You Can't Open an LLM With a Screwdriver''}: The De-Democratization of Software} 

\author{Zixuan Feng}
\affiliation{%
  \institution{Virginia Commonwealth University}
  \city{Richmond}
  \state{Virginia}
  \country{USA}
}
\email{fengz3@vcu.edu}

\author{Italo Santos}
\authornote{Co-first author}
\affiliation{%
  \institution{University of Hawai\textquotesingle i at M\=anoa}
  \city{Honolulu}
  \state{Hawaii}
  \country{USA}
}
\email{isantos3@hawaii.edu}

\author{Kostadin Damevski}
\affiliation{%
  \institution{Virginia Commonwealth University}
  \city{Richmond}
  \state{Virginia}
  \country{USA}
}
\email{kdamevski@vcu.edu}

\author{Anita Sarma}
\affiliation{%
  \institution{Oregon State University}
  \city{Corvallis}
  \state{Oregon}
  \country{USA}
}
\email{anita.sarma@oregonstate.edu}

\begin{abstract}

Claims that generative AI will soon write all of the code have led to predictions that programming is nearing its end. In this vision paper, we argue against this assumption that broader access to code generation necessarily democratizes software development, i.e., everyone can code but we have to distinguish between \emph{access} and \emph{control}: by access, we mean the ability of more people, including non-experts and less-experienced developers, to generate code-like artifacts with AI; by control, we mean the capacity to inspect, evaluate, integrate, maintain, and govern those artifacts as dependable software. While AI may broaden access to code production, control may become more concentrated among those who own or understand the code, software practices, infrastructure, evaluation practices, and deployment pipelines. 

Grounded in an expert panel, our vision paper argues that AI does not eliminate software engineering expertise but shifts where that expertise becomes most critical. The locus of software engineering expertise is shifting toward \textit{intent specification}: orchestrating and governing AI behavior, evaluating software behavior, and integrating software systems. We conclude this paper by identifying research opportunities for education, tools, and policy that can help the software engineering community respond to the AI era with greater agency, accountability, and adaptability.

\end{abstract}

\maketitle


\section{Introduction}
 
For a decade, ``learn to code'' was marketed as a golden ticket for economic mobility, creating thousands of job positions and generating billions in economic value~\cite{jorgenson2003growth, brynjolfsson2021productivity}. However, it has now been abruptly inverted by claims from leading AI laboratories that we are months away from AI doing what software engineers do~\cite{haider2026anthropic}. A premise once controversial is now becoming mainstream: in the near future, almost all code will be written by AI~\cite{bergman2026aiwritingstartupcode, cfr2025amodei}. The appealing promise is \emph{democratization} of software development: AI appears to make software development almost accessible to anyone, including domain experts, end users, and people with little or no programming background.

In this paper, we contest the democratization narrative. We argue that the democratization conflates \emph{access} with \emph{control}. Access refers to the ability to produce software artifacts, a capability that AI broadens~\cite{Kazemitabaar_2023}. Control refers not simply to producing code, but to knowing what is being built, teaching AI how to build~\cite{welsh2022end}, evaluating whether it behaves as intended, integrating it into existing systems, deploying it responsibly, and taking accountability for its consequences. While AI expands access to code generation, it may also widen the gap between generating software artifacts and exercising responsible oversight over them, while unsettling traditional ownership and accountability rules around AI-generated artifacts~\cite{fagan2026autonomous}.

Therefore, we argue that ``democratization'' is revealed as an \emph{illusion}. The impact of AI is not the elimination of the software engineer or the obsolescence of computer science education; it is a shift in professional roles, scarce skills, and what we should teach the next generation in the software industry.

This is not the first end-of-programming moment. Since FORTRAN's \emph{automatic coding} era in the 1950s, successive waves of abstraction, from model-driven engineering and program synthesis to low-code/no-code platforms, have promised to move software development away from line-by-line coding and toward higher-level specification~\cite{Fortran1957, backus1978history,france2007model, yan2021impacts}. Today's AI coding systems revive that promise in a more powerful form. Welsh's ``The End of Programming'' exemplifies the claim that classical programming, i.e., writing code by hand, is nearing its end because of AI~\cite{welsh2022end}.

Yet the history of abstraction suggests a different lesson: abstraction changes where the difficulty resides, not eliminating it. As Dijkstra warned, programming remains the use of a formal symbolism and continues to require care and accuracy~\cite{dijkstra2005foolishness}; as Brooks argued, there is no silver bullet~\cite{brooks1987no} to instantly fix a complex problem; and as Naur argued, programming is a form of \emph{theory building}, where the human contribution is not merely code but the mental model of why a system is the way it is~\cite{naur1985programming}.

Our vision is that programming is not disappearing because of AI. Instead, as in earlier waves of technological change, AI is shifting the scarce skills of software engineering from writing code line by line to specifying intent, evaluating behavior, integrating generated artifacts, maintaining systems, guiding AI-generated code, and governing the infrastructures that produce it. Grounded in an expert panel discussion, this vision paper identifies where the field’s uncertainty lies and turns that uncertainty into a set of research opportunities for the software engineering ecosystem.

\label{sec:intro}

\section{Methodology}
To ground our vision discussion, we adopted an expert-informed qualitative approach inspired by expert elicitation~\cite{morgan2014use,hemming2018practical}. We qualitatively analyzed an expert panel discussion on AI in software as the empirical basis for our vision to identify recurring concerns, emerging opportunities, unresolved tensions, and future research directions.

\begin{figure}[!ht]
    \centering
    \includegraphics[width=0.95\linewidth]{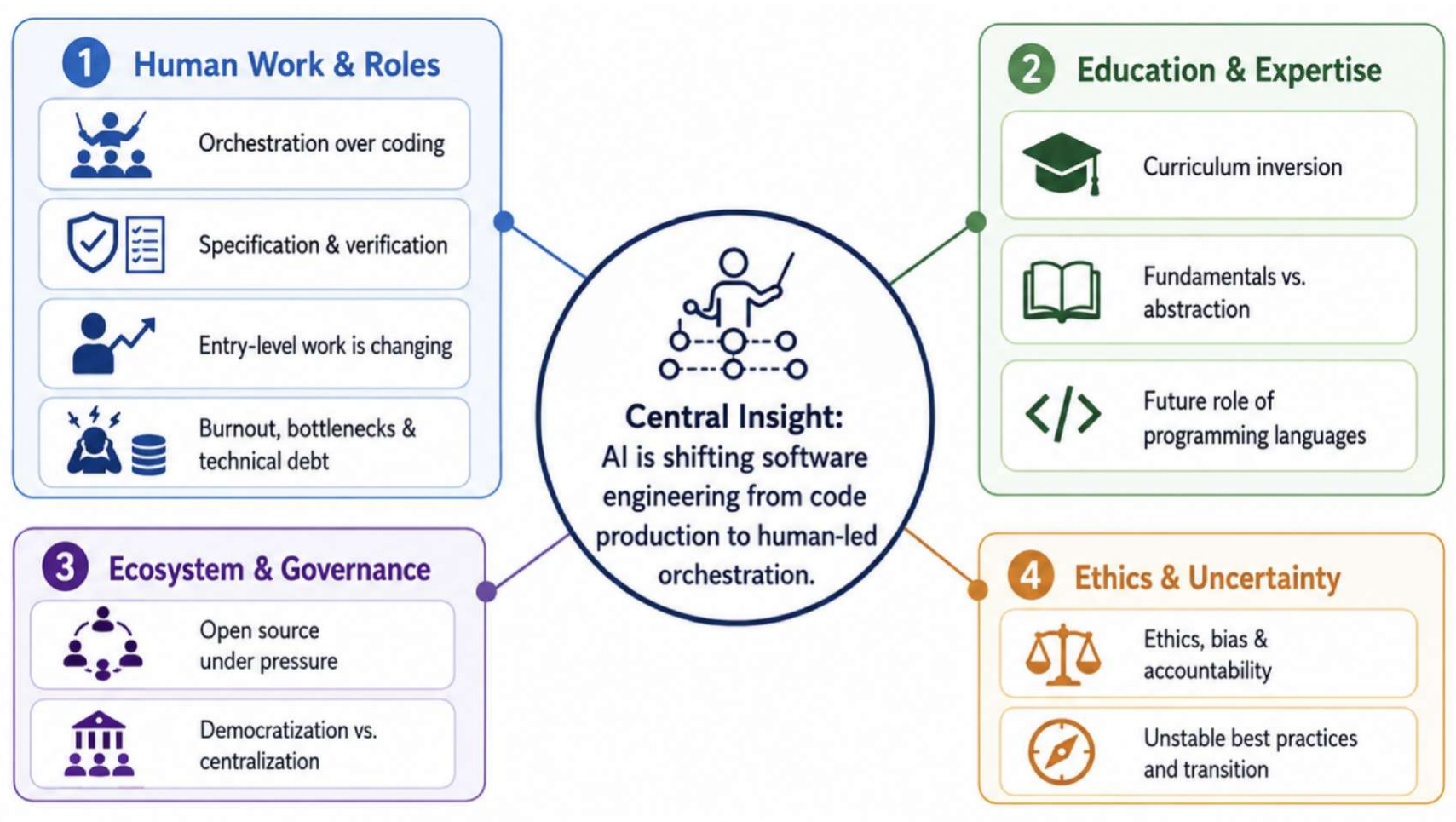}
    \caption{Thematic map of human role in AI-mediated software engineering}
    \label{fig:overview}
    \vspace{-3mm}
\end{figure}

\textbf{The expert panel.}  The panel ``Future of Software'' was conducted in person at Oregon State University as part of the \emph{Oregon AI Research Workshop} on April 17, 2026 \footnote{https://engineering.oregonstate.edu/all-stories/future-software-when-ai-writes-code-what-do-humans-do}.

The panel was moderated by a senior computer science, AI, and robotics faculty member. The panelists consisted of academic and industry experts, including NVIDIA’s Director of Deep Learning Systems Software, a senior software-engineering education faculty, two AI researchers working on verification, reinforcement learning, and multi-agent systems, and an empirical software engineering researcher focused on open-source sustainability.

\textbf{The discussion.} 
The panel discussion covered a set of questions about the future of AI-mediated software engineering, including the human role in software development; how computer science education may need to respond; what software itself may become; the future role of programming languages; the future of open source software; and the risks, surprises, and unresolved questions raised by these shifts.

The discussion began from a widely discussed premise: in the near future, most code may be written by AI. Rather than treating this as the end of software engineering, the panel used this premise to discuss what remains essentially human in software development. The discussion focused on the specification and formalization of intent, verification, judgments of fitness for purpose, system-level orchestration, integration, and domain expertise.

The panel then turned to the consequences of this shift for software work and education. Panelists discussed how career ladders and entry-level roles may change when new hires increasingly supervise agents rather than write code by hand; whether computer science curricula should be ``inverted'' to foreground systems thinking before low-level programming; and how programming languages may evolve if human legibility becomes less central than agent efficiency, verification, and token cost.

The discussion then broadened from professional practice to the software ecosystem. Panelists considered how AI-generated contributions may strain open-source capacity, how accelerated code generation may amplify technical and security debt, and how model bias and the concentration of AI capability among a small number of companies raise ethical and governance concerns. In the final Q\&A, audience questions pushed these issues further, asking whether AI-mediated software development democratizes software production or instead shifts control toward more closed, centralized infrastructures.

\textbf{Analysis.} We transcribed the session and conducted a thematic analysis~\cite{braun2006using}. The first two authors began by familiarizing themselves with the transcript. They then independently analyzed the transcript to identify recurring claims about the changing human role in AI-mediated software engineering. Two researchers met through weekly meetings to compare interpretations, discuss points of divergence, and refine themes through negotiated agreement.

During the analysis, we focused especially on moments of disagreement and contestation: points where panelists disagreed, where audience members pushed back, or where the same AI capability appeared to create both promise and harm. These unresolved tensions motivated our vision to identify where software engineering research is most needed. As \emph{research happens where tension happens}~\cite{feng2025domains}.

The resulting themes were then reviewed against the full transcript. As shown in Figure~\ref{fig:overview}; our analysis developed four higher-level thematic categories: \emph{Human Work and Roles}, \emph{Education and Expertise}, \emph{Ecosystem and Governance}, and \emph{Ethics and Uncertainty}. These categories capture the panel's core insight: AI is shifting software engineering from direct code production toward human-led orchestration.

\textbf{Scope, limitations, and positionality.} We acknowledge that this paper draws on a single panel discussion. The panel brought together researchers and practitioners whose expertise is around AI and software. The panelists' backgrounds shaped what the discussion made visible: it foregrounded questions of human expertise, governance, education, accountability, and production constraints, while leaving other perspectives, such as those of students, early-career developers, and policymakers, less represented. We therefore treat the panel as a grounding lens for motivating our vision, and we make no claim of statistical generalizability.

Our interpretation is also shaped by our positions as an author team of four academic faculty members across three institutions. Our collective expertise lies at the intersection of software engineering, AI-mediated software development, software engineering education, open-source sustainability, empirical software engineering, and socio-technical systems. The authors bring complementary perspectives on how software is produced, learned, evaluated, maintained, and governed. We write from the standpoint of academic software engineering researchers, and our interpretation should not be read as representing the full range of industrial, educational, policy, or global perspectives on AI-mediated software engineering. We use our collective expertise to discuss a vision and identify research opportunities for the software engineering community.

\label{sec:theory}

\section{Results}

Grounded by our qualitative analysis, we frame our vision around the four themes shown in Figure~\ref{fig:overview}. We argue that \emph{AI is shifting software engineering from code production to human-led orchestration: while AI appears to democratize software engineering by lowering the barrier to producing code, it can also reproduce new forms of dependency, opacity, and exclusion.}

\textbf{1. Human Work \& Roles: From coding to orchestration.} Panelists repeatedly described the future engineer as an \emph{orchestrator}: an ``executive chef'' coordinating agents, modules, tests, and integration. The human capability is not simply writing code, but specifying intent \textit{``we are the arbiters of intent''}, and verifying fitness-for-purpose: deciding whether a generated, often unread, artifact actually serves its deployed context. Prior empirical research with software professionals similarly shows that developers use LLMs as assistive tools while retaining responsibility for reviewing, adapting, testing, and validating generated code~\cite{santosmodel,testingscam,arewetesting}.

This also changes the impact of the entry-level work. The panel resisted the claim that junior roles simply disappear; junior work moves up the stack toward reviewing designs, clarifying intent, authoring tests, decomposing tasks for agents, and integrating generated components. At the same time, the new human work and roles create a new workload problem. AI-generated contributions can arrive at a speed and volume that existing review, integration, and accountability practices were never designed to handle. \textit{``AI is a cart with rocket fuel: it can help a project reach its goal faster, but it can also drive it off the cliff ``a hundred times faster''}, as technical debt at machine speed.

\textbf{2. Education \& Expertise: Curriculum inversion and the return of fundamentals.} If human work is moving from coding to orchestration, the panel debated whether computing education should move in the same direction. One position was to \emph{invert the curriculum}: teach systems thinking, problem decomposition, software quality, specification, verification, and architecture early, while moving low-level programming to later stages. In this view, students should be trained for the work AI makes scarce: framing problems, supervising agents, and judging whether generated systems are reliable enough for production.

But this proposal immediately exposed a counter-tension. Panelists warned that abstraction without a conventional computer science foundation becomes fragile: students still need to know what the abstractions are abstracting \emph{of}. Prior research in open source projects shows that learners and newcomers may struggle to identify the skills required for software tasks and benefit from structured, personalized support that helps them evaluate and complete complex engineering activities~\cite{hitandmisses,greatpower,ossdoorway}. Programming is less like typing code line by line; but the student needs sufficient conceptual understanding to debug, evaluate, and recognize when a tool has gone wrong. The future of programming languages may become less important as everyday development tools, but add more focus on substrates for verification, agent communication, new architectures, and disciplined expressions of intent. Thus, AI does not make expertise disappear; it changes the educational question from ``Should students still learn to code?'' to ``What minimum grounding lets students responsibly supervise systems they may no longer fully hand-write?''

\textbf{3. Ecosystem \& Governance: Open source under pressure.} The panel widened the discussion to the software ecosystem. Open source became a sample stress case: if AI can generate code faster than humans can review it, then open communities become vulnerable to contribution floods. Panelists described their open source team maintainers receiving large volumes of AI-assisted pull requests and issues, many of which may look plausible but still require human attention to evaluate. This pressure exposes a governance problem in open source that is damaging the software engineering ecosystem. With a large amount of AI-generated low-quality contributions, communities may respond defensively, such as closing issues \cite{itsfoss2026curl, claburn2025curl_ai_slop_bug_bounty}.

Another tension lies between democratization and centralization. AI appears to broaden access by enabling more people to build and contribute software, yet the means of production increasingly depend on opaque, corporate-controlled models and platforms. As a result, the same tools that make software creation more accessible may also make software ecosystems more dependent, filtered, and governed through new forms of gatekeeping.

\begin{table*}[!ht]
\centering
\caption{From cross-cutting tensions to research opportunities.}
\label{tab:tensions}
\scriptsize
\renewcommand{\arraystretch}{1.22}
\setlength{\tabcolsep}{4pt}
\begin{tabularx}{\textwidth}{|
>{\raggedright\arraybackslash}p{0.16\textwidth}|
>{\raggedright\arraybackslash}X|
>{\raggedright\arraybackslash}X|
>{\raggedright\arraybackslash}X|}
\hline
\textbf{Cross-cutting tension} &
\textbf{Problem} &
\textbf{Stakes} &
\textbf{Research opportunity} \\
\hline
\textbf{Productivity $\leftrightarrow$ Burnout} &
AI increases the volume and speed of code, tests, pull requests, and entire software production cycle, but human review, integration, and decision-making do not scale at the same pace. &
The risk is that productivity gains become cognitive overload, bottlenecks, and technical debt at machine speed. &
Design sociotechnical workflows that scale human oversight: agent triage, specification-level diffing, AI-assisted review, and coordination mechanisms that keep humans in control without forcing them to operate at machine speed. \\
\hline
\textbf{Abstraction $\leftrightarrow$ Fundamentals} &
AI enables higher-level interaction with software, but students and engineers still need to understand what the abstractions are built on. &
Without grounding, abstraction becomes fragile when generated systems fail or require debugging. &
Identify the irreducible mental model needed to supervise AI-generated systems, and design curricula that teach systems thinking, programming fundamentals, verification, and debugging without relying only on syntax-heavy struggle. \\
\hline
\textbf{Democratization $\leftrightarrow$ Dependency} &
AI lowers the barrier to producing software, but the deeper means of production increasingly depend on opaque models, cloud platforms, APIs, and corporate-controlled infrastructure. &
Access may broaden at the surface while control narrows underneath. &
Develop re-democratized AI infrastructure: auditable models, local or edge inference, transparent provenance, open ecosystems, and governance mechanisms that reduce dependency on closed platforms. \\
\hline
\textbf{Automation $\leftrightarrow$ Accountability} &
AI can generate, test, and help verify software, but humans remain responsible for security, fairness, reliability, and fitness-for-purpose. &
Accountability becomes harder to locate when artifacts are opaque, non-deterministic, and only partially inspected. &
Create assurance methods that connect AI-generated artifacts back to human intent and deployment context, including traceability, provenance, bias auditing, security review, and post-generation validation. \\
\hline
\end{tabularx}
\end{table*}

\textbf{4. Ethics \& Uncertainty: Automation without settled accountability.} Finally, the panel emphasized that AI-mediated software engineering is still moving through an unstable transition. AI can generate code, propose fixes, and even assist with review, but it does not remove human responsibility for the consequences of software. Panelists raised concerns about bias in training data, culturally sensitive applications, security vulnerabilities, and the risk of over-trusting systems whose reasoning and failure modes remain difficult to inspect. In this sense, the human role is not only technical but also ethical: engineers must decide whether AI-generated outputs are safe, fair, appropriate, and accountable in the contexts in which they will be deployed.

Yet the panel also made clear that best practices remain unsettled. Software industries are still experimenting with when to trust AI, when to require disclosure, how much human review is enough, and how to govern workflows that change faster than policies can stabilize.

\section{Research Opportunities}

The optimistic reading (anyone can build) and the pessimistic reading (the software engineer is obsolete) are both partial. We now convert the panel's tensions into research opportunities. Table~\ref{tab:tensions} maps each tension to a research opportunity.

\textbf{1. Productivity vs.\ burnout: scaling output without scaling exhaustion.} AI makes developers faster, but it also increases the overall volume of software production beyond what existing practices can absorb. Engineers must inspect outputs, switch across agent-generated tasks, resolve integration failures, and prevent technical debt from accumulating. When software output is produced at machine speed, productivity gains can quickly turn into burnout, bottlenecks, and fragile systems \cite{afroz2026fast, feng2025gains}. Future work should examine how agent hierarchies, review triage, specification-level diffing, AI-assisted review, and workflow designs redistribute attention, accountability, and control in AI-mediated software development.

\textbf{2. Abstraction vs.\ fundamentals: knowing what the abstraction hides.} AI enables students to work at higher levels of abstraction before they have acquired lower-level programming experience. On one side, this is promising because it allows curricula to foreground systems thinking, decomposition, specification, verification, and software quality earlier. On the other hand, it is also risky because abstraction becomes fragile when students do not understand what the abstraction is built on. When AI-generated systems fail, students and engineers still need to debug, evaluate, and reason across layers. Future research should investigate what forms of foundational knowledge remain necessary for supervising AI-generated software, how much programming and systems grounding is enough, and how curricula can balance early abstraction with the durable fundamentals needed to diagnose and govern AI-mediated systems.

\textbf{3. Democratization vs.\ dependency: broad access built on narrow control.} AI appears to democratize software engineering because more people can now produce working programs. The problem is that this democratization often happens at the surface of software production, while the deeper means of production become more centralized. Models, platforms, data centers, APIs, pricing structures, and safety filters are controlled by a small number of organizations. This matters because users may gain the ability to build while losing the ability to inspect, repair, own, or contest the infrastructure on which building depends. Future research should therefore examine how software engineering can preserve AI’s accessibility gains while reducing dependency on opaque platforms. This includes studying open and auditable models, local and edge inference, transparent provenance, public-interest AI infrastructure, and governance mechanisms that make AI-mediated software production more inspectable, contestable, and accountable.

\textbf{4. Automation vs.\ accountability: assuring more than correctness.} The problem is not only that AI may generate incorrect code, but that it can generate software whose accountability is difficult to trace. Existing assurance practices often focus on whether an artifact passes tests or satisfies a specification, yet many consequential requirements live outside the artifact itself: domain constraints, user harms, security assumptions, bias, maintainability, organizational values, and legal responsibility. This matters because AI-generated systems can be technically plausible while still violating the intent, context, or obligations of the people who deploy them. Future research should therefore investigate assurance methods that preserve accountability across the AI-mediated lifecycle, including how intent is specified, how generated artifacts are reviewed, how responsibility is assigned, and how harms are detected after deployment.

\label{sec:survey}

\section{Conclusion}

AI lets more people produce plausible software. But \textit{``LLM is not something that I can open up with a screwdriver and inspect''}: a model you did not build, cannot inspect, cannot repair, must be rented by the token, and which can refuse to act.  We argue that today is the least democratized time in software's history. Democratization is only at the level of outputs and illusory at the level of control, while the bar for good software continues to rise. The capability to produce, evaluate, and govern AI-generated software remains concentrated in a handful of firms and their data centers.

There are many tensions that the panel experts have not reached agreement on and don't have solutions; these tensions are valuable to treat as challenges for the software engineering research community with the breadth, expertise, and responsibility to shape the next era of software engineering. Future work should investigate these challenges empirically, develop educational models and theories for AI-mediated software engineering, and design tools and governance structures that preserve human agency, accountability, and dependable software practice.

\bibliographystyle{ACM-Reference-Format}
\bibliography{bib}
\end{document}